\documentclass[a4paper,11pt]{article}
\usepackage{graphicx,natbib}
\usepackage{lineno,hyperref,amsmath,amssymb}
\usepackage{verbatim}
\usepackage[hmargin=2.5cm]{geometry}

\renewcommand{\d}{\:\mathrm{d}}
\newcommand{\del}{\partial}
\newcommand{\tld}[1]{\ensuremath{\tilde{#1}}}
\newcommand{\x}{\tld{x}}
\newcommand{\y}{\tld{y}}
\renewcommand{\t}{\tld{t}}
\newcommand{\mat}[1]{\ensuremath{\mathbf{#1}}}

\DeclareMathOperator{\arcsinh}{arcsinh}

\title{Shear localisation controls the dynamics of earthquakes}

\author{Fabian Barras$^1$ and Nicolas Brantut$^2$\\
  $^1$ The Njord Centre, University of Oslo, Oslo, Norway\\
 $^{2}$ Department of Earth Sciences, University College London, London, UK}

\date{\ }

\begin{document}

\maketitle

\begin{abstract}
Earthquakes are produced by the propagation of rapid slip along tectonic faults. The propagation dynamics is governed by a balance between elastic stored energy in the surrounding rock, and dissipated energy at the propagating tip of the slipping patch. Energy dissipation is dictated by the mechanical behaviour of the fault, which is itself the result of feedbacks between thermo-hydro-mechanical processes acting at the mm to sub-mm scale. Here, we numerically simulate shear ruptures using a dual scale approach, allowing us to couple a sub-mm description of inner fault processes and km-scale elastodynamics, and show that the sudden localisation of shear strain within a shear zone leads to the emergence of classical cracks driven by a constant fracture energy. The fracture energy associated to strain localisation is substantially smaller than that predicted assuming uniform shearing. We show the existence of a unique scaling law between the localised shearing width and the rupture speed. Our results indicate that earthquakes are likely to be systematically associated to extreme strain localisation.
\end{abstract}

\section{Introduction}

Earthquake sources correspond to slip events dynamically propagating along faults. At crustal scale, faults can be viewed as two-dimensional surfaces, across which the displacement field is discontinuous. However, geological and experimental observations show that ``slip'' across faults is the result of shear deformation across narrow layers of highly comminuted, transformed or partially melted rocks. In the shallow continental crust, fault core materials are often made of fine-grained siliclastic and clay gouges, with a porosity filled with pressurised water \citep[e.g.][]{scholz88,rice06}. The dynamics of ruptures in crustal faults is controlled by the rheology of these water-saturated fault gouges.

During earthquakes, faults slide at elevated slip rates of the order of metres per second, which leads to dramatic weakening of fault gouge materials \citep[][chap. 2]{scholz19}. In dry materials, weakening is most likely controlled by the local temperature rise arising from dissipation of frictional work, combined with thermally activated rheology of the rock-forming minerals \citep[e.g][]{rice06,beeler08,proctor14,depaola15,yao15,pozzi21,harbord21}. In the presence of fluids, an additional weakening mechanism is expected, due to the differential thermal expansion of the pore fluids and the solid pore space: upon heating, the fluid pressure rises, effective normal stress decreases and the frictional strength drops. This so-called ``thermal pressurisation'' mechanism, initially proposed by \citet{sibson75} as a temperature-limiting process in wet rocks, has been shown to produce realistic predictions for thermal evolution and energy dissipation during earthquakes \citep[e.g.][]{rice06,viesca15}, and is a potential candidate to explain some of the complexity observed in natural earthquakes \citep[e.g.][]{noda13} and the operation of plate boundary faults at low ambient stress \citep[e.g.][]{noda09,lambert21}.

The thickness of the actively deforming zone determines the shear heating rate and how easily fluids and heat diffuse away from the fault plane, and thus has a tremendous influence on the resulting rupture dynamics \citep{andrews02,noda09,noda10,viesca15,lambert21}. While geological and experimental observations can be used to constrain the thickness of actively deforming fault gouge material, the range of acceptable values spans more than 3 orders of magnitude, from fractions of millimetres to centimetres \citep{rice06}, and it is one of the key unknown that limits our ability to determine the efficiency of thermal weakening mechanisms in nature.

The influence of shear zone width on earthquake propagation is further complicated by the fact that this parameter is likely evolving during seismic slip: strain localisation is expected to be part of the fault weakening process. Several mechanisms might be responsible for strain localisation during earthquake slip, including granular rearrangements and grain size reduction \citep[e.g.,][]{mair08,hermundstadt10}, shear heating coupled to thermal weakening \citep[e.g.,][]{braeck07}, thermal pressurisation \citep[e.g.,][]{sulem11,rice14,platt14} and thermal decomposition \citep[e.g.,][]{veveakis12,platt15}. In all cases, the strain localisation process is associated to a rapid reduction in shear strength, and we therefore expect strain localisation to exert a strong control on the overall dynamics of rupture.

Here, we demonstrate and quantify how strain localisation impacts rupture dynamics: we run dynamic rupture simulations and compute the fracture energy associated with the localisation process, and find a relationship between the rupture speed and the degree of strain localisation within the fault gouge. We use the case of thermal pressurisation as a representative thermal weakening process that is compatible with seismological observations \citep{rice06,viesca15,lambert21}, and is known to spontaneously lead to strain localisation \citep{rice14,platt14}. We argue that the interplay between rupture dynamics and strain localisation analysed here applies to most thermal weakening processes in rocks and engineering materials. 

\begin{figure}
  \centering
  \includegraphics{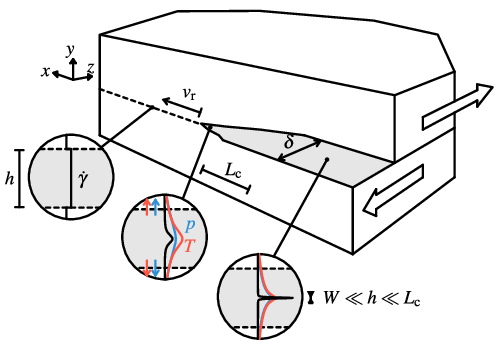}
  \caption{Schematic of the dual scale setup governing the propagation of localised shear bands. The dynamic rupture extends over large (kilometric) scales along the fault ($x-z$ plane), whereas the frictional strength is determined by solving a coupled diffusion problem, such as thermal pressurisation in this paper, where strain rate spontaneously evolves over submillimetre scales across the fault gouge (in the $y$ direction).}
  \label{fig:system_setup}
\end{figure}

\section{Shear localisation and faulting}

As a general starting point for our analysis, let us consider slip on geological fault as the deformation distributed across a narrow pre-existing shear zone (e.g., a fault gouge made of unconsolidated grains), and examine the conditions under which shear strain can be further localised within this pre-existing zone. Let us assume that the shear stress $\tau$ across the shear zone is function of a set of variables that includes the shear strain rate $\dot{\gamma}$ and another diffusive quantity $\vartheta$: 
\begin{linenomath}
  \begin{equation}
  \tau\sim f(\dot{\gamma},\vartheta),
  \label{equ:shearevol}
\end{equation}
\end{linenomath}
and that the rate of work produced by shearing acts as a source term in the diffusion of $\vartheta$:
\begin{linenomath}
  \begin{equation}
  \dot\vartheta = \beta\tau\dot\gamma + \alpha\nabla^2\vartheta,
  \label{equ:difftheta}
\end{equation}
\end{linenomath}
where $\nabla^2$ denotes the Laplace operator, $\alpha$ is a diffusivity and $\beta$ is analogous to the Taylor-Quiney coefficient \citep{taylor34b} for the cases when $\vartheta$ corresponds to temperature. In relation to the first condition, one can define
\begin{linenomath}
  \begin{equation}
  g'(\dot{\gamma},\vartheta) = \frac{\partial f}{\partial\dot{\gamma}},
  \label{equ:fprime}
\end{equation}
\end{linenomath}
which describes the rate-dependent rheology of the material. Natural examples include viscous creep of rocks at elevated temperature, granular material rheology \citep[e.g.][]{jop2006} and rate-and-state friction. Similarly, one can define
\begin{linenomath}
  \begin{equation}
  h'(\dot{\gamma},\vartheta) = \frac{\partial f}{\partial\vartheta},
  \label{equ:gprime}
\end{equation}
\end{linenomath}
to describe the effect of $\vartheta$ on the material rheology. In practice, this diffusive quantity $\vartheta$ will often correspond to temperature and $h'$ describe thermal weakening effects. $\vartheta$ could also correspond to fluid pressure in a porous material whose strength is reduced by an increase in pore fluid pressure (following the concept of effective stress discussed later in Equation \ref{equ:shear_rheol}). It can also account for the combined effect of pressure and temperature as in the case of thermal pressurisation that will be discussed later in this manuscript. If conditions \eqref{equ:shearevol} and \eqref{equ:difftheta} are met, a linear stability analysis (detailed in Appendix \ref{sec:applinpert}) demonstrates that uniform shearing at a given time $t=t_0$ becomes unstable if:
\begin{linenomath}
  \begin{equation}
\frac{h'_0}{g'_0} < 0, 
\label{equ:locacrit}
\end{equation}
\end{linenomath}
with $\lbrace f_0, g'_0, h'_0\rbrace=\lbrace f, g',h'\rbrace|_{t=t_0}$. Moreover, the analysis also shows that only perturbation wavelengths $\lambda$ greater than a critical wavelength are unstable:
\begin{linenomath}
  \begin{equation}
    \lambda > 2\pi\sqrt{-\frac{\alpha}{\beta f_0}\frac{g'_0}{h'_0}} \equiv \lambda_{\rm c},
    \label{equ:critwavelength}
  \end{equation}
\end{linenomath}
which indicates that such instability leads to the localisation of shear strain down to some thickness $W_{\rm loc}\sim\lambda_{\rm c}/2$. Remarkably, this type of localisation instability can also arise within rate-strengthening materials ($g'_0>0$) providing that $h'_0<0$, as it is often the case with thermal weakening mechanisms. As a result, shear flow concentrates over a thickness much smaller than the initial width of the shear zone and leads to a substantive drop of the associated shear stress. Examples of this strain localisation instability have been described by \cite{rice14,platt14} in the context of thermal pressurisation in crustal faults, as well as \cite{bai82} in the context of adiabatic shear banding in metals. In this work, we quantitatively investigate how strain localisation across the shear zone drives the rapid acceleration of slip and the propagation of rupture front along the fault plane during earthquakes.

\section{Model}

Here, we analyse the process of thermal pressurisation, which has been shown to be a realistic dynamic weakening mechanism \citep[e.g.][]{rice06,viesca15,lambert21} and that undergoes the localisation instability outlined above \citep{rice14,platt14}. 

In this case, the diffusive variable $\vartheta$ corresponds to pore fluid pressure $p$ that affects the effective normal stress $(\sigma_{\rm n}-p)$ in the shear zone and, thereby, its shear strength together with a rate-dependent friction coefficient:
\begin{linenomath}
  \begin{equation}
    \tau(\dot{\gamma},p) = f_{\rm rsf}(\dot{\gamma})(\sigma_{\rm n}-p).
    \label{equ:shear_rheol}
  \end{equation}
\end{linenomath}
In Equation \eqref{equ:shear_rheol}, we adopt the rate-strengthening rheology $f_{\rm rsf}(\dot{\gamma})$ of \cite{platt14} detailed in Appendix \ref{sec:app:TP}. Moreover, fluid diffusion across the shear zone is governed by a coupled system of thermal  and hydraulic diffusion equations (see Equations \ref{equ:diffTP} in Appendix). This thermo-hydraulic coupling is caused by the different compressibilities and thermal expansivities of the solid matrix and the pore fluid, and describes the change in pore pressure produced by local temperature variations in the gouge.

Exploring the interplay between strain localisation and rupture dynamics is a challenging dual-scale problem: it requires solving for heat and fluid diffusion at the scale of the fault core (from millimeters to centimeters in natural fault zones) together with the elastodynamics governing the propagation of the earthquake rupture along the fault (elastic waves moving at kilometers per second in crustal rocks). We follow \citet{noda09} and take advantage of the separation of scale to solve thermal pressurisation only across the fault (along the $y$ axis in Figure \ref{fig:system_setup}). In this one-dimensional configuration, \cite{platt14} found numerically that shear localisation, under imposed constant slip rate $V$ across the gouge, stabilises to a finite width that is well approximated by
\begin{linenomath}
  \begin{equation}
    W_{\rm rsf}(V) \simeq 6.9\frac{A\rho c}{\Lambda f_{\rm c}}\frac{(\sqrt{\alpha_{\rm hy}}+\sqrt{\alpha_{\rm th}})^2}{V(f_{\rm c}+2A)},
    \label{equ:wloc}
  \end{equation}
\end{linenomath}
where $\alpha_{\rm hy, th}$ correspond respectively to the hydraulic and thermal diffusivities, $\rho c$ is the heat capacity of the gouge material,  $\Lambda$ is the thermal pressurisation parameter that describes the change of pore fluid pressure caused by an increase in temperature in the gouge. The characteristic shear strength
\begin{linenomath}
  \begin{equation}
\tau_{\rm c}=f_{\rm c}(\sigma_{\rm n} - p_0)=f_{\rm rsf}(V/h)(\sigma_{\rm n} - p_0)
\label{equ:tau0}
\end{equation}
\end{linenomath}
is a function of the initial uniform strain rate $\dot{\gamma}=V/h$ and background pore pressure $p_0$. In Equation (\ref{equ:wloc}), the constant $A$ corresponds to the rate-strengthening coefficient that describes the "direct" response of the shear zone to a change in strain rate similar to standard rate-and-state models (see Equation (\ref{equ:friction_law}) in Appendix). Once the localisation width $W_{\rm rsf}$ is much smaller that the diffusion length scale, thermal pressurisation can be well approximated by a \textit{slip-on-a-plane} solution assuming $\dot{\gamma}(y)=V\bar\delta(y)$, with $\bar\delta$ being the Dirac delta distribution. In this situation, the shear stress across the shear zone is only function of the imposed slip rate $V$ and the accumulated slip $\delta = Vt$ \citep{rice06,mase87}:
\begin{linenomath}
  \begin{equation}
     \tau_{\rm sp}(\delta; V) = \tau_{\rm c}\exp\Big(\frac{\delta}{L^*}\Big)\mathrm{erfc}\Big(\sqrt{\frac{\delta}{L^*}}\Big),
     \label{equ:sliponaplane}
   \end{equation}
 \end{linenomath}
 where
 \begin{linenomath}
   \begin{equation}
     L^*(V) = \frac{4}{f^2_{\rm c}}\Big(\frac{\rho c}{\Lambda}\Big)^2\frac{(\sqrt{\alpha_{\rm hy}}+\sqrt{\alpha_{\rm th}})^2}{V}.
     \label{equ:Lstar}
   \end{equation}
 \end{linenomath}

During earthquakes, slip rate across the shear zone is far from being constant and evolves rapidly near the tip of the propagating rupture. Next, we aim to analyse the coupling between strain localisation, slip acceleration and rupture dynamics in a simple faulting geometry that is sufficient to capture its key physical aspects. We consider a planar fault within an infinite linear elastic medium sliding in anti-plane shear (mode-III). In this configuration shown in Figure \ref{fig:system_setup}, the shear traction at the fault $\tau(x,t)$ can be related to the slip rate across the fault $V(x,t)$ and, thereby, the strain rate in the shear zone, following 
\begin{linenomath}
  \begin{equation}
\tau(x,t) = \tau_{\rm b} - \frac{\mu}{2c_{\rm s}}V(x,t) + \phi(x,t) = \tau_{\rm b} - \frac{\mu}{2c_{\rm s}}\int_{-h/2}^{h/2}\dot{\gamma}(x,y,t)\d y + \phi(x,t).
\label{equ:elastodyn}
\end{equation}
\end{linenomath}
In the equation above, $\mu$ is the shear modulus, $c_{\rm s}$ the shear wave speed of the linear elastic medium surrounding the shear zone, $\phi$ is the non-local dynamic contribution accounting for the history of slip along the interface, and $\tau_{\rm b}$ represents the far-field background stress. Equation (\ref{equ:elastodyn}) allows us to define the characteristic seismic slip rate $V_{\rm c}$ and associated uniform strain rate $\dot{\gamma}_{\rm c}$ as
\begin{linenomath}
  \begin{equation}
    \dot{\gamma}_{\rm c} = \frac{V_{\rm c}}{h} = \frac{2c_{\rm s}\tau_{\rm b}}{h\mu},
  \end{equation}
\end{linenomath}
which are used in the remainder of the paper together with the related characteristic shear strength $\tau_{\rm c}$ (\ref{equ:tau0}). The elastodynamic equation \eqref{equ:elastodyn} couples the strain rate in the shear zone $\dot{\gamma}$ to the shear stress $\tau$ and allows us to implement a dual-scale coupled numerical scheme that solves the rupture elastodynamics along the shear zone together with pressure and temperature diffusion across the shear zone. The details of our coupled numerical scheme are given in Appendix \ref{sec:app_numerical}.

\section{Results}

\begin{figure}
  \centering
  \includegraphics{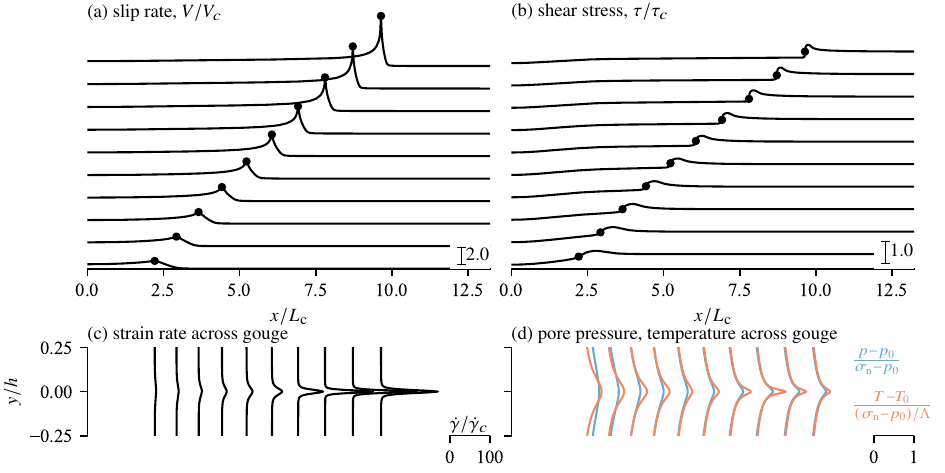}
  \caption{Dynamic rupture driven by shear localisation simulated with the coupled model. The top panels a) and b) respectively present snapshots at different times of the longitudinal profile of slip rate and shear stress during which the rupture accelerates from sixty to about ninety percent of the shear wave velocity. Note that the simulated domain is symmetric with respect to the nucleation position $x=0$ such that another rupture tip moves toward the negative positions. The bottom panels c) and d) present the profile of strain rate $\dot{\gamma}$, pressure $p$ and temperature $T$ at the successive positions of the rupture tip highlighted by black dots in panel a) and b). See Appendix \ref{sec:app_dimensionless} and Table \ref{Tab:dimensionless} for further details on the dimensional analysis behind this coupled problem and the dimensionless scaling used to plot the data in the different panels.}
  \label{fig:typical_sim}
\end{figure}

\begin{figure}
  \centering
  \includegraphics{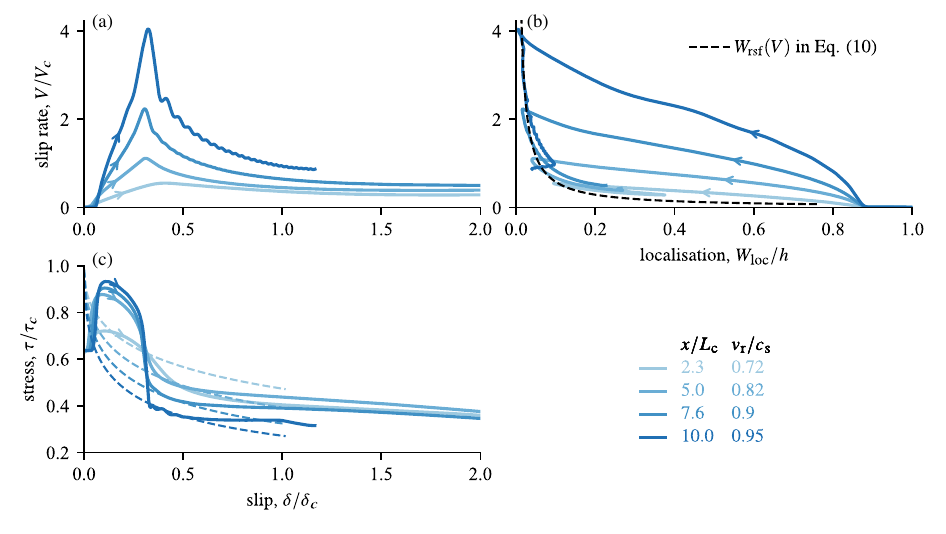}
  \caption{Time evolution of the elastic variables at different locations along the interface during the dynamic rupture shown in Figure \ref{fig:typical_sim}. Line colors relate to the positions along the interface $x/L_{\rm c}$ and the associated propagation speeds of the rupture $v_{\rm r}/c_{\rm s}$, whereas the arrows point to the direction of forward time evolution. Slip rate (a) and shear stress (c) versus slip revealing how the peak slip rate is associated to abrupt stress drop and arise at the same amount of cumulated slip $\delta_{\rm loc}\approx0.3\delta_{\rm c}$. (b) Slip rate versus width of strain rate localisation $W_{\rm loc}$ measured from the $\dot{\gamma}(y)$ profiles following the procedure shown in Figure 3 of \cite{platt14}. The different post-peak delocalisation trajectories collapse along a single prediction given in Equation (\ref{equ:wloc}). The dashed lines in panel (c) correspond to the prediction $\tau_{\rm sp}(\delta; V_{\rm tip})$ and gives a good prediction of the residual shear stress reached after strain localisation. The slip rate at the rupture tip $V_{\rm tip}$ is approximated by $V$ at the mid-time between the peaks in shear stress and in slip rate. (A more precise definition of the tip position is discussed and computed later in the context of Figure \ref{fig:lefm_fit}.)}

  \label{fig:space_snapshot}
\end{figure}

In our simulations, the shear zone is initially creeping at aseismic slip velocity and, at time $t=0$, failure is nucleated by rising the pore pressure near the center of the fault $x=0$ (further details of nucleation procedure and parameter values are given in Appendix \ref{sec:app_numerical}). Initially, acceleration of slip is mostly concentrated in the nucleation region, followed by a rapid lateral rupture propagation whereby the slip rate increases in an expanding region beyond the initial nucleation patch, concomitantly with a shear stress drop linked with thermal pressurisation of pore fluids inside the gouge and intense strain localisation (Figure \ref{fig:typical_sim}). Rupture acceleration coincides with larger slip velocities and stress drop at the tip (Figure \ref{fig:space_snapshot}a-c) and more intense localisation of shear deformation across the gouge where up to four orders of magnitude larger strain rate concentrates on less than five percent of the thickness of the shear zone (Figure \ref{fig:space_snapshot}b). Interestingly, the peak slip rate and drop of shear stress measured at different positions along the fault arise for the same characteristic slip value $\delta_{\rm loc}$ and coincides with intense strain localisation. The observed value of $\delta_{\rm loc}$ is identical to the one reported from one-dimensional simulation under imposed velocity and is in the order of magnitude of $\delta_{\rm c}$ (see Figures 9 and 10 of \cite{platt14}). Remarkably, this observation enables us to apply the one-dimensional theory discussed in the previous section to derive predictions of the shear zone dynamics after strain localisation. For instance, the slip-on-a-plane solution described in Equation (\ref{equ:sliponaplane}) can be used to capture the magnitude of the residual shear stress reached immediately after strain localisation $\tau_{\rm res}\approx\tau_{\rm sp}(\delta=\delta_{\rm c};V=V_{\rm tip})$, with $V_{\rm tip}$ being the slip rate observed at the rupture tip (see Figure \ref{fig:space_snapshot}c and related caption). Moreover, once the localisation instability arises, the thickness of actively strained material at various positions along the interface collapses on a single $W_{\rm loc}(V)$ curve, which follows the prediction given in Equation (\ref{equ:wloc}).

\subsection{Rupture dynamics driven by shear localisation}

\begin{figure}
  \centering
  \includegraphics{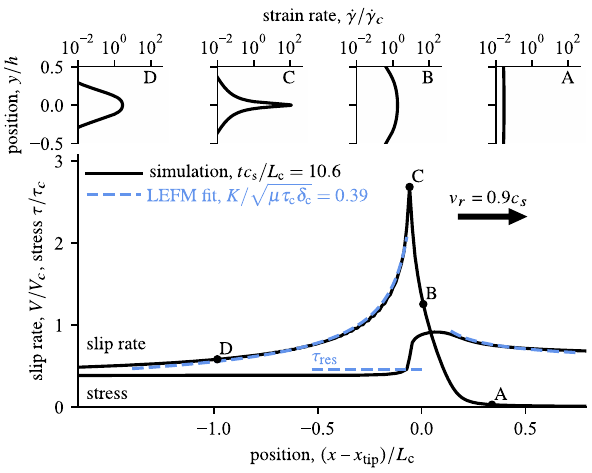}
  \caption{Snapshot near the tip of the propagating rupture shown in Figure \ref{fig:typical_sim}. Bottom panel presents the spacial evolution of the shear stress and slip rate, which are simultaneously fitted by the fracture mechanics prediction shown by the dashed blue curve. (See the main text for details on the fitting procedure). Top panels show the strain rate profile across the shear zone observed at the instants A,B,C and D corresponding to the black dots in the bottom panel.}
  \label{fig:lefm_fit}
\end{figure}

Next, we quantitatively demonstrate how strain localisation is the driving mechanism of the propagating rupture. To do so, we analyze snapshots of the propagating rupture and the near-tip evolution of the macroscopic and microscopic mechanical variables (Figure \ref{fig:lefm_fit}). Ahead of the propagating tip (point A), the shear zone is creeping with uniform shear strain rate. As the rupture approaches, the strain rate builds up uniformly across the gouge (point B) until the localisation instability arises (point C) together with a rapid increase in macroscopic slip rate $V$ and abrupt drop of shear stress $\tau$. In the wake of the rupture (point D), the profile of strain rate across the gouge progressively delocalises, following the decay of the macroscopic slip rate given by the prediction $W_{\rm rsf}(V)$ shown in Figure \ref{fig:space_snapshot}b. The near-tip evolution of $V$ and $\tau$ is reminiscent to the singular solutions at the tip of a dynamic fracture \citep{freund90}. Defining $\alpha^2_{\rm s} = 1-v^2_{\rm r} /c^2_{\rm s}$, the analogy to linear elastic fracture mechanics (LEFM) can be quantitatively tested by rescaling the slip rate and stress according to
\begin{linenomath}
  \begin{equation}
    V \frac{\mu\alpha_{\rm s}}{2v_{\rm r}} = \tau-\tau_{\rm res} = \Delta\tau = \frac{K}{\sqrt{2\pi(x-x_{\rm tip})}}
    \label{equ:v_peak_pred}
  \end{equation}
\end{linenomath}
and fitting the dynamic fracture solution following the procedure of \citet{barras20}. The stress intensity factor $K$, residual stress $\tau_{\rm res}$ and position of the tip $x_{\rm tip}$ are the free parameters that are fitted simultaneously to match the near-tip decrease of $V$ behind the rupture tip and increase of $\tau$ ahead of the rupture tip.

\begin{figure}
  \centering
  \includegraphics{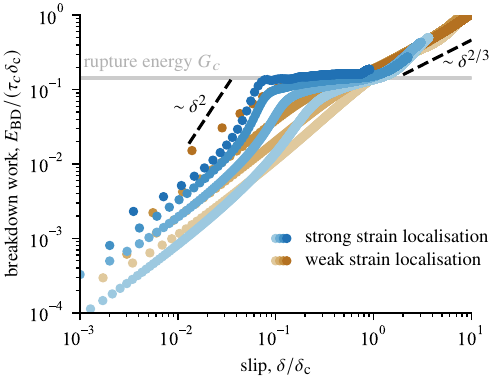}
  \caption{Breakdown energy integrated from the $\tau$ versus $\delta$ profiles at different positions along the fault, for the simulation shown in Figure \ref{fig:space_snapshot} (labelled ``strong strain localisation'', blue dots) and for another simulation with the same parameters but larger hydraulic diffusivity (labelled ``weak strain localisation'', brown dots). The rupture simulated with larger hydraulic diffusivity shows very weak strain localisation. (Additional plots of the dynamics observed during that simulation are given in Figures \ref{fig:app_snapshot_undrained} and \ref{fig:app_lefmfit_undrained} of the Appendix). The rapid loss of stress caused by strain localisation creates an horizontal plateau whose associated magnitude is well predicted by the rupture energy inverted from the dynamic fracture fit shown in Figure \ref{fig:lefm_fit} and highlighted here by the horizontal gray line.}
  \label{fig:stress_EBD}
\end{figure}

The good agreement with dynamic fracture solution (dashed blue curves in Figure \ref{fig:lefm_fit}) confirms the crack-like nature of the simulated rupture process near the tip of the slipping patch. Such agreement allows us to use the inverted value of $K$ and invoke the crack-tip energy balance to compute the rupture energy
\begin{linenomath}
  \begin{equation}
    G_{\rm c} = \frac{K^2}{2\mu\alpha_{\rm s}},
  \end{equation}
\end{linenomath}
which corresponds to the part of dissipated energy that governs the propagation of the rupture. In seismology, extracting the fracture energy of natural earthquakes still eludes the resolution of seismic inversions, such that the \textit{breakdown work} is often used as a proxy for $G_{\rm c}$ and integrates the excess of work on top of residual friction \citep{tinti05,cocco23}. For systems where frictional weakening does not necessarily reach a well-defined residual value, the breakdown work is defined as \citep{abercrombie05}:
\begin{linenomath}
  \begin{equation}
    E_{\rm BD}(\delta) = \int_0^\delta\Big(\tau(\delta')-\tau(\delta)\Big)\d\delta'.
    \label{equ:EBD}
  \end{equation}
\end{linenomath}
In our numerical simulations, the integration of $E_{\rm BD}$ at different locations along the interface reveals a clear plateau over an order of magnitude in slip (Figure \ref{fig:stress_EBD}), which indicates the portion of $E_{\rm BD}$ that corresponds to $G_{\rm c}$ following \citet{brener21}. Remarkably, we can then quantitatively verify that the two independent estimates of the rupture energy (from the near-tip singularity and the integration of the breakdown work) are in excellent agreement (gray horizontal line in Figure \ref{fig:stress_EBD}) as another proof of the crack-like nature of the rupture dynamics. Furthermore, the observed plateau in $E_{\rm BD}$ is clearly associated to the rapid stress drop caused by localisation instability (see $\tau(\delta)$ profile in Figure \ref{fig:space_snapshot}c) and confirms that rapid strain localisation is the driving mechanism of the propagating rupture. In addition, the magnitude of $G_{\rm c}$ associated to strain localisation is more than five times smaller than that expected from uniform shearing under adiabatic undrained conditions ($\sim\tau_{\rm c}\delta_{\rm c}$). 

The interplay between strain localisation and rupture dynamics can be further established by relating the thinnest localisation width observed at a given location along the interface to the local speed of the rupture (Figure \ref{fig:wloc_vs_speed}). Following the behavior reported in Figure \ref{fig:space_snapshot}a, the dynamic stress drop caused by the localisation instability can be well estimated by the slip-on-a-plane solution (\ref{equ:sliponaplane}). Together with the elastodynamic relation (\ref{equ:v_peak_pred}), it relates the slip rate near the rupture tip $V_{\rm tip}$ to the rupture speed $v_{\rm r}$, which can further be combined to the solution $W_{\rm rsf}(V_{\rm tip})$ of Equation (\ref{equ:wloc}):
\begin{linenomath}
  \begin{equation}
\begin{cases}
   V_{\rm tip} = \displaystyle \frac{2v_{\rm r}}{\mu\alpha_{\rm s}}\Big(\tau_{\rm c}-\tau_{\rm sp}(\delta_{\rm c}; V_{\rm tip})\Big),\\
   W_{\rm loc} = W_{\rm rsf}(V_{\rm tip}).
\end{cases}
\label{equ:Wloc_versus_vr}
\end{equation}
\end{linenomath}
The implicit relation above provides a universal relationship between $v_{\rm r}$ and the degree of localisation $W_{\rm loc}$ observed at the scale of the gouge. Its good agreement with different simulations data (Figure \ref{fig:wloc_vs_speed}) unveil the dynamical feedback loop at play between strain localisation and rupture dynamics: the dynamic stress drop caused by strain localisation drives the acceleration of slip, which further amplifies the dynamic stress drop and promotes rupture acceleration. 

\section{Discussion}

\begin{figure}
  \centering
  \includegraphics{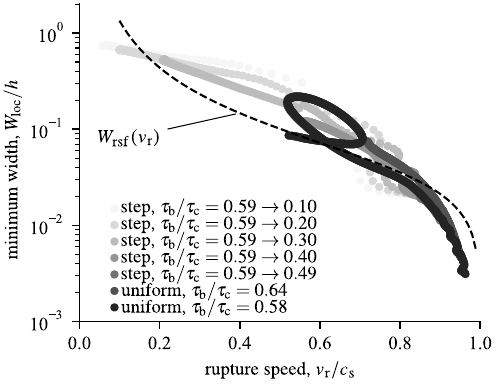}
  \caption{Minimum strain rate localisation width $W_{\rm loc}$ versus instantaneous rupture speed $v_{\rm r}$ computed during rupture propagation for several simulations using the same parameters but different background stresses. Heterogeneous simulations with steps in background stress were conducted with an initial plateau of $\tau_\mathrm{b}/\tau_\mathrm{c}=0.59$ around the nucleation zone, and a sharp drop down to a smaller value at position $x/L_\mathrm{c}=\pm11.52$ away from the center of the nucleation region. In heterogeneous simulations, rupture speeds may vary nonmonotonically during propagation, initially increasing and subsequently decreasing when encountering a large downstep in stress. Regardless of the details of the dynamics, the relationship between peak localised width and rupture speed is well approximated by the theoretical prediction proposed in Equation (\ref{equ:Wloc_versus_vr}).}

  \label{fig:wloc_vs_speed}
\end{figure}

Our simulations demonstrate that strain localisation produces a loss of stress-bearing capacity of shear zones that can create and sustain earthquake rupture. The abrupt drop of shear stress produces an accelerating crack-like rupture in agreement with the predictions of LEFM. Notably, the rupture is driven by a well-defined fracture energy that corresponds to the edge-localised dissipation during the localisation process. 

Such a behaviour is in contrast with that of ruptures driven by thermal pressurisation without internal strain localisation \citep{viesca15}, for which breakdown work uniformly increases with increasing slip, i.e., without a well-defined residual strength at any point within the propagating rupture. Similarly, the integrated breakdown work for simulated ruptures that feature weak strain localisation lack a well-defined edge-localised rupture energy (Figure \ref{fig:stress_EBD}). In this case, the rupture tip singularity significantly deviates from fracture mechanics predictions (\ref{equ:v_peak_pred}) and the rupture dynamics is no longer governed by local near-tip energy balance \citep{brener21}, as further discussed in Appendix \ref{sec:app_unconv}. Far from the rupture tip, our simulations show further shear weakening driven by the diffusion process (\ref{equ:difftheta}) that continues at a slower pace as strain delocalises, and we approach again the slip-on-a-plane asymptotics described by \citet{viesca15}.

Strain localisation within a preexisting gouge material is strongly correlated to the dynamics of fault slip, and specifically to the rupture speed (Figure \ref{fig:wloc_vs_speed}). The degree of strain localisation increases with increasing rupture speed, with a narrowing of the deformed, heated and pressurised region, approaching 1/1000th of the initial shear zone width. Despite the complexity of the problem, quantitative estimates can be obtained by a simple analytical approximation (Equation \ref{equ:wloc}) adapted from \citet{platt14}, so that the original predictions for peak localised width listed in \citet{rice14,platt14} still apply. Ideally, we could use the relationship between width and rupture speed depicted in Figure \ref{fig:wloc_vs_speed} to interpret the localisation features observed in the geological record in terms of rupture dynamics. However, strain localisation in rocks is not exclusively associated with dynamic ruptures and fast slip rates \citep[e.g.][]{evans95}, and only careful micro- and nano-structural studies can be relied upon to determine the seismic nature of geological structures, notably via detection of features characteristic of frictional heating \citep{rowe15}. Keeping this caveat in mind, our results highlight that the degree of strain localisation may be used as a complementary indicator of seismic slip: indeed, simulations leading to dynamic ruptures are always associated with strong localisation, with typical width in the sub-millimeter range \citep[see also][]{daub08}.

In this paper, we chose thermal pressurisation as the driving mechanism for localisation and implemented a numerical scheme coupling small-scale diffusive processes across the shear zone to long-range elastodynamic coupling along the shear zone. Our results can however be generalized to other type of localisation instability arising within shear zones where (1) the shear stress in the shear zone is function of a set of variables that includes the shear strain rate $\dot{\gamma}$ and another diffusive quantity $\vartheta$ (Equation \ref{equ:shearevol}), and (2) the rate of work produced by shearing acts as a source term in the diffusion of $\vartheta$ (Equation \ref{equ:difftheta}). Importantly, shear localisation can produce and sustain rupture in shear zones having a rate-strengthening rheology ($g'_0>0$) often interpreted as a token of stability and aseismic slip. 

If the conditions (\ref{equ:locacrit}) and (\ref{equ:critwavelength}) are fulfilled, a localisation instability can develop and lead to an abrupt drop of shear stress, which leads to the emergence of a well-defined edge-localised fracture energy and LEFM-like rupture. Far from the tip, any diffusion-driven weakening leads to $E_{\rm BD}\sim \delta^{2/3}$ at large slip \citep{brantut17b}. Therefore, the behavior summarized in Figure \ref{fig:stress_EBD} is expected to arise for any type of localisation-driven rupture, including those where the rheology is controlled by temperature, such as superplasticity \citep[e.g.][]{green15,depaola15,pozzi21,harbord21}. Indeed, simulations of high speed deformation in metals, which are also rate-hardening and temperature-sensitive, tend to exhibit similar characteristics, with the emergence of a localisation-driven dissipation at the edge of propagating shear bands \citep{bonnet-lebouvier02}.

Our work demonstrates how localisation instabilities arising across a creeping shear zone create an abrupt drop of shear stress that promotes the propagation of classical dynamic ruptures over large distances along the shear zone. Whether frictional systems are governed by classical fracture mechanics or by nonlinear friction is an important and debated question in geophysics \citep[e.g.][]{svetlizky19,lambert21,paglialunga22,cocco23}. Strain localisation is an abrupt structural weakening mechanism that provides a clear separation between the cohesive zone and the interior of the slipping patch, hence justifying the small-scale yielding hypothesis. However, the relative simplicity of the rupture tip behaviour does not preclude any complexity of the overall rupture style. Away from the rupture tip, thermal and hydraulic diffusion and strain delocalisation maintain a slow decay of the shear stress, which is prone to impact how earthquake ruptures stop \citep{paglialunga22}.

\bibliography{localbib}
\clearpage 
\appendix
\section{Linear perturbation analysis \label{sec:applinpert}}

Let us consider a shear zone as in Figure \ref{fig:system_setup} initially creeping under imposed shear stress $\tau_0 = f_0$ and uniform strain rate $\dot{\gamma}_0(t)$ and field $\vartheta_0(t)$ conditions across the shear zone following the constitutive Equations \eqref{equ:shearevol} and \eqref{equ:difftheta}:
\begin{linenomath}
  \begin{equation}
    \partial_t\vartheta_0 = \beta f_0\dot{\gamma}_0,
    \label{equ:app_uniform_conds}
  \end{equation}
\end{linenomath}
with $\partial_t$ denoting a partial time derivative. At $t=t_0$, small perturbations to the uniform configuration are introduced such that the evolution of the three variables of interest can be written as
\begin{linenomath}
  \begin{equation}
    \left\lbrace\begin{array}{c}
         \tau(y,t) = f_0 + \tau_1(y,t)\\
         \dot{\gamma}(y,t) = \dot{\gamma}_0(t) + \dot{\gamma}_1(y,t)\\
         \vartheta(y,t) = \vartheta_0(t) + \vartheta_1(y,t)
    \end{array}\right. ,
    \label{equ:lin_pert}
  \end{equation}
\end{linenomath}
with $\lbrace\tau_1/f_0,\dot{\gamma}_1/\dot{\gamma}_0,\vartheta_1/\vartheta_0\rbrace\ll 1$. Using the definitions of $g'_0$ and $h'_0$ given in the main text (\ref{equ:fprime}),(\ref{equ:gprime}), the constitutive law governing the shear zone rheology can be expanded as
\begin{linenomath}
  \begin{equation}
f_0 + \tau_1 = f_0 + g'_0\dot{\gamma}_1 + h'_0\vartheta_1 + \mathcal{O}(\dot{\gamma}^2_1,\vartheta^2_1),
\end{equation}
\end{linenomath}
and further simplified as
\begin{linenomath}
  \begin{equation}
    \tau_1 = g'_0\dot{\gamma}_1 + h'_0\vartheta_1,
    \label{equ:app_tau1}
  \end{equation}
\end{linenomath}
neglecting the higher-order terms. The diffusion equation can be similarly linearised 
\begin{linenomath}
  \begin{equation}
    \partial_t\vartheta_0 + \partial_t\vartheta_1 = \beta(f_0\dot{\gamma}_0 +f_0\dot{\gamma}_1 + \tau_1\dot{\gamma}_0) + \alpha\partial^2_y\vartheta_1
    \label{equ:diffperturb}
  \end{equation}
\end{linenomath}
and simplified further using the uniform shear solution (\ref{equ:app_uniform_conds}). The remaining terms write
\begin{linenomath}
  \begin{equation}
    \partial_t\hat{\vartheta}_1 = \beta(f_0\dot{\hat{\gamma}}_1 + \hat{\tau}_1\dot{\gamma}_0) - 4\pi^2 k^2\alpha\hat{\vartheta}_1,
    \label{equ:diffperturbspec}
  \end{equation}
\end{linenomath}
using the following spectral decomposition
\begin{linenomath}
  \begin{equation}
    \lbrace\hat{\tau}_1,\dot{\hat{\gamma}}_1,\hat{\vartheta}_1\rbrace(k,t) = \int^{\infty}_{-\infty}\lbrace\tau_1,\dot\gamma_1,\vartheta_1\rbrace(y,t) e^{-2\pi iy}\d y.
  \end{equation}
\end{linenomath}
From the conservation of momentum and neglecting inertia through the thickness of the shear zone ($\partial_y \tau = 0$), one has $\hat{\tau}_1 = 0$ such that Equation (\ref{equ:app_tau1}) writes
\begin{linenomath}
  \begin{equation}
     \dot{\hat{\gamma}}_1 = - \frac{h'_0}{g_0'}\hat{\vartheta}_1.
   \end{equation}
 \end{linenomath}
Equation (\ref{equ:diffperturbspec}) becomes
\begin{linenomath}
  \begin{equation}
    \partial_t\hat{\vartheta}_1 = -\hat{\vartheta}_1\Big(\beta f_0\frac{h_0'}{g_0'} + 4\pi^2 k^2\alpha\Big)
  \end{equation}
\end{linenomath}
and allows the following solution
\begin{linenomath}
  \begin{equation}
    \hat{\vartheta}_1 = \hat{\vartheta}_{0}\exp{\Big((t-t_0)(-4\pi^2 k^2\alpha-\beta f_0\frac{h_0'}{g_0'})\Big)},
  \end{equation}
\end{linenomath}
with $\hat{\vartheta}_{0}$ being the initial value of $\hat{\vartheta}_1$ at $t=t_0$
Consequently, uniform creeping across the shear zone is stable if
\begin{linenomath}
  \begin{equation}
    k^2 > -\frac{\beta f_0}{4\pi^2\alpha}\frac{h'_0}{g'_0},
  \end{equation}
\end{linenomath}
which leads to the conditions (\ref{equ:locacrit}) and (\ref{equ:critwavelength}) in the main text.

\section{Model}
\subsection{Elastodynamics (longitudinal problem)}

The longitudinal problem is solved assuming that the shear zone lays between two semi-infinite half-spaces. The elastodynamics is then solved using a boundary integral formulation that relates the traction $\sigma_{yi}(x,y=\pm h/2,t)$ acting within the shear zone to the respective displacements of each surfaces of the surrounding elastic wall rock $u_i(x,y=\pm h/2,t)$. We consider homogeneous elastic properties and anti-plane shear conditions, such that the shear traction in the shear zone $\tau(x,t)=\sigma_{yz}$ can be related to the differential slip $\delta(x,t) =u_z(x,y=h/2,t)-u_z(x,y=-h/2,t)$ and slip rate $V(x,t) =\dot{u}_z(x,y=h/2,t)-\dot{u}_z(x,y=-h/2,t)$ following Equation (\ref{equ:elastodyn}), in which $\phi$ accounts for the non-local elastodynamic interactions and is evaluated in the Fourier domain as:

\begin{linenomath}
  \begin{equation}
\Phi(k,t) = -\frac{1}{2}\mu|k|D(k,t) + \frac{1}{2}\mu|k|\int_{-\infty}^t W_{\rm III}\Big(|k|c_{\rm s}(t-t')\Big)\dot{D}(k,t')\d t',
\label{equ:convol}
\end{equation}
\end{linenomath}
with $\Phi(k,t)$, $D(k,t)$ and $\dot{D}(k,t)$ being respectively the Fourier transform pairs of $\phi(x,t)$, $\delta(x,t)$ and $V(x,t)$. The mode III elastodynamic kernel $W_{\rm III}$ is defined in \citet{morrissey97} as a function of the Bessel function of the first kind $J_1$:
\begin{linenomath}
  \begin{equation}
    W_{\rm III}(T) = \int_T^\infty \frac{J_1(\zeta)}{\zeta}d\zeta.
    \label{equ:kernel}
  \end{equation}
\end{linenomath}

\subsection{Thermal pressurisation (transverse problem) \label{sec:app:TP}}
The transverse problem accounts for the thermo-hydro-mechanical shear response of a fluid-saturated fault gouge. Taking advantage of the time and length scale separation between the $x$ and $y$ directions existing in the problem of interest, heat and fluid flows through the granular material are only solved along the $y$ direction. Next, the assumption of fluid flow dominated by viscosity (Darcy's law with hydraulic diffusivity $\alpha_{\rm hy}$) and heat transfer by conduction (Fourier's law with thermal diffusivity $\alpha_{\rm th}$) are invoked to obtained a coupled set of diffusion equations that describes change of temperature $T$ caused by frictional shear and the associated change in fluid pressure $p$ due to thermal expansion of the porous medium. The \textit{thermal pressurisation} equations read \citep[e.g.][]{rice06}:
\begin{linenomath}
  \begin{equation}
\begin{aligned}
    \frac{\del T}{\del t} &= \frac{\tau\dot{\gamma}}{\rho c} + \alpha_{\rm th} \frac{\del^2 T}{\del y^2}, \\
    \frac{\del p}{\del t} &= \Lambda\frac{\del T}{\del t} + \alpha_{\rm hy} \frac{\del^2 p}{\del y^2}.
\end{aligned}.
\label{equ:diffTP}
\end{equation}
\end{linenomath}
In the equations above, $\rho$ and $c$ are respectively the density and specific heat capacity of the gouge and $\Lambda$ is a mean-field parameter that describes the change of pore fluid pressure caused by an increase in temperature in the gouge. $\tau$ and $\dot{\gamma}$ are respectively the shear stress and strain rate in the gouge. Following \citet{rice14,platt14}, we assume that the gouge follows a weak rate-hardening behavior: 
\begin{linenomath}
  \begin{equation}
\tau = f_{\rm rsf}(\dot{\gamma})(\sigma_{\rm n}-p) = \Big(f_0 + A\ln(\dot{\gamma}/\dot{\gamma}_0)\Big)(\sigma_{\rm n}-p),
\label{equ:friction_law}
\end{equation}
\end{linenomath}
defined by the friction parameters $f_0$, $A$ and $\dot{\gamma}_0$ and the normal stress $\sigma_{\rm n}$. By analogy to rate-and-state friction, $A$ controls the "direct" strengthening response of the gouge whereas the "long term" evolution of the shear stress is given by the thermo-hydro-mechanical response of the gouge through the evolution of $p(y,t)$. 

\subsection{Coupling conditions}

Throughout the rupture, the separation of scales between the longitudinal and transverse direction allows us to neglect 1) fluid and heat flows in the $x$ direction and 2) inertial effects through the thickness of the gouge. Consequently, the shear stress is invariant across the gouge and corresponds to 
\begin{linenomath}
  \begin{equation}
    \tau\Big(x,-h/2\leq y\leq h/2,t\Big) \equiv \tau(x,t).
    \label{equ:cstshearstress}
  \end{equation}
\end{linenomath}
Conversely, the strain rate varies across the gouge and is related to the macroscopic slip velocity via the following integration:
\begin{linenomath}
  \begin{equation}
    \int_{-h/2}^{h/2} \dot{\gamma} \d y = \int_{-h/2}^{h/2} 
    \frac{\partial\dot{u}}{\partial y} \d y = \dot{u}(x,y=h/2,t) - \dot{u}(x,y=-h/2,t) \equiv V(x,t).
    \label{equ:vtostrainrate}
  \end{equation}
\end{linenomath}
The stress (\ref{equ:cstshearstress}) and kinematic (\ref{equ:vtostrainrate}) conditions above allow for coupling the longitudinal and transverse problems.

\section{Dimensional analysis \label{sec:app_dimensionless}}
Defining $V_{\rm c}=2c_{\rm s}\tau_{\rm b}/\mu$, $t_{\rm c}=\delta_{\rm c}/V_{\rm c}$ and $L_{\rm c}=c_{\rm s}t_{\rm c}$, the elastodynamic equation (\ref{equ:elastodyn}) can be rewritten in dimensionless form as
\begin{linenomath}
  \begin{equation}
    \tld\tau(\x,\t) = 1 + \tld V(\x,\t) + \tld\phi(\x,\t),
    \label{equ:adim_elastodyn}
  \end{equation}
\end{linenomath}
with $\tld\tau = \tau/\tau_{\rm b}$, $\x=x/L_{\rm c}$, $\t=t/t_{\rm c}$, $\tld V=V/V_{\rm c}$ and
\begin{linenomath}
  \begin{equation}
\Phi(\tld k,\t) = -\tld D(\tld k,\t) + |\tld k|\int_{-\infty}^{\t} W_{\rm III}\Big(|\tld k|(\t-\t')\Big)\dot{\tld D}(\tld k,\t')\d \t'.
\end{equation}
\end{linenomath}
The characteristic slip distance $\delta_{\rm c}$ will be set later by the rheology of the interface. Defining $\dot{\gamma}_{\rm c}=V_{\rm c}/h$ and $f_{\rm c}=f_{\rm rsf}(\dot{\gamma}_{\rm c})$, the effective friction relationship can be rewritten as
\begin{linenomath}
  \begin{equation}
\begin{aligned}
    \tld\tau &= \Big(1 + \frac{A}{f_{\rm c}}\ln(\dot{\gamma}/\dot{\gamma}_{\rm c})\Big)\Big(1-\frac{p-p_0}{\sigma_{\rm n}-p_0}\Big)\frac{f_{\rm c}(\sigma_{\rm n}-p_0)}{\tau_{\rm b}}\\
    &\equiv\Big(1+\tld z^{-1}\ln(\dot{\tld\gamma})\Big)\Big(1-\tld p\Big)\tld\eta^{-1}
\end{aligned}    
\end{equation}
\end{linenomath}
In the equation above, $\tld z+\ln(\dot{\tld\gamma})$ is further approximated as $\arcsinh(e^{\tld z}\dot{\tld\gamma}/2)$ in order to rewrite the constitutive relationship as
\begin{linenomath}
  \begin{equation}
    \tld\tau = \frac{1-\tld p}{\tld z\tld\eta}\arcsinh\Big(\frac{\dot{\tld\gamma}}{2}e^{\tld z}\Big),
    \label{equ:adimfrictionlaw}
  \end{equation}
\end{linenomath}
which has the advantage of regularizing the stick-to-slip transition as $\dot{\tld\gamma}\to0$. In dimensionless form, the kinematic condition (\ref{equ:vtostrainrate}) writes then 
\begin{linenomath}
  \begin{equation}
    \tld{V}(\x,\t) = \int_{-1/2}^{1/2}\dot{\tld\gamma}(\x, \y, \t)\;\d\y = \int_{-1/2}^{1/2}2e^{-\tld z}\sinh\Big(\frac{\tld z\tld\eta\tld\tau(\x, \t)}{1-\tld p(\x, \y, \t)}\Big)\;\d\y.
    \label{equ:vfromtau}
  \end{equation}
\end{linenomath}
The set of diffusion equations governing thermal pressurisation \eqref{equ:diffTP} can then be written in the following dimensionless form
\begin{linenomath}
  \begin{equation}
\begin{aligned}
    \frac{\del \tld T}{\del \t} &= \tld\eta\tld\tau\dot{\tld\gamma} + \tld \alpha_{\rm th}\frac{\del^2 \tld T}{\del \y^2}\\
    \frac{\del \tld p}{\del \t} &= \frac{\del \tld{T}}{\del \t} + \tld \alpha_{\rm hy}\frac{\del^2 \tld{p}}{\del \y^2}
\end{aligned}
\label{equ:adimTP}
\end{equation}
\end{linenomath}
and provides the definition of the characteristic slip
\begin{linenomath}
  \begin{equation}
    \delta_{\rm c} = \frac{\rho ch}{\Lambda f_{\rm c}}.
  \end{equation}
\end{linenomath}
Remarkably, in this dimensionless framework summarized in Table \ref{Tab:dimensionless}, the behavior of the coupled system is controlled only by the following four dimensionless quantities:
\begin{linenomath}
  \begin{equation}
\left\lbrace
\begin{aligned}
\tld \alpha_{\rm th} &= \alpha_{\rm th}\frac{t_{\rm c}}{h^2} = \alpha_{\rm th}\frac{\mu\rho c}{2c_{\rm s}\tau_{\rm b} f_{\rm c}\Lambda h},\\
\tld \alpha_{\rm hy} &= \alpha_{\rm hy}\frac{t_{\rm c}}{h^2} = \alpha_{\rm hy}\frac{\mu\rho c}{2c_{\rm s}\tau_{\rm b} f_{\rm c}\Lambda h},\\
\tld z &= \frac{f_{\rm c}}{A}, \\
\tld\eta &= \frac{\tau_{\rm b}}{f_{\rm c}(\sigma_{\rm n}-p_0)}, \\
\end{aligned}
\right.
\end{equation}
\end{linenomath}
recalling that $f_{\rm c}=f_{\rm rsf}\Big(\dot{\gamma}=\frac{2c_{\rm s}\tau_{\rm b}}{h\mu}\Big)$. The reference set of parameters used for the simulations are representative to a fault zone at 7~km depth \citep{platt14} and corresponds to $\tld \alpha_{\rm th}=0.0061$, $\tld \alpha_{\rm hy}= 0.075$, $\tld z = 15$ and $\tld\eta = 0.64$.

\begin{table}
\everymath{\displaystyle}
     \begin{center}
     \begin{tabular}{l c c } \hline   
        Quantities    & Variables & Definition \\ \hline
        Pressure  & $\tld p$   & $(p-p_0)/(\sigma-p_0)$ \\ 
        Temperature  & $\tld T$   & $\Lambda(T-T_0)/(\sigma-p_0)$ \\
        Distance across gouge & $\y$ & $y/h$ \\
        Shear stress & $\tld\tau$ & $\tau/\tau_{\rm b}$ \\
        Slip velocity  & $\tld v=V/V_{\rm c}$   & $v\times\mu/(2c_{\rm s}\tau_{\rm b})$ \\
        Slip & $\tld\delta=\delta/\delta_{\rm c}$ & $\delta \times \Lambda f_{\rm c}/(\rho ch)$ \\
        Distance along fault & $\x=x/L_{\rm c}$ & $x \times (2\tau_{\rm b})/(\mu\delta_{\rm c})$ \\
        Time  & $\t=t/t_{\rm c}$ & $t\times (2c_{\rm s}\tau_{\rm b})/(\mu\delta_{\rm c})$ \\
        Strain rate & $\dot{\tld\gamma}=\dot{\gamma}/\dot{\gamma}_{\rm c}$ & $\dot{\gamma}\times(\mu h)/(2c_{\rm s}\tau_{\rm b})$ \\ \hline
    \end{tabular}
   \caption{Table of dimensionless variables}\label{Tab:dimensionless}
 \end{center}
\end{table}

\section{Numerical methods \label{sec:app_numerical}}
\subsection{System of diffusion equations}
The two diffusion Equations (\ref{equ:adimTP}) solved in the transverse direction can be generically written as
\begin{linenomath}
  \begin{equation}
    \frac{\del \tld \Psi}{\del \t} = \tld S + \tld D\frac{\del^2 \tld \Psi}{\del \y^2},
  \end{equation}
\end{linenomath}
and are numerically integrated using Crank-Nicholson method with an explicit source term:
\begin{linenomath}
\begin{align*}
&\frac{\tld \Psi_j^{m+1}-\tld \Psi_j^{m}}{\Delta\tld t} = \tld S^m_j \\ &+ \frac{\tld D}{\Delta\tld y_{j+1}+\Delta\tld y_j}\Big(\frac{\tld \Psi_{j+1}^{m+1}-\tld \Psi_{j}^{m+1}}{\Delta\tld y_{j+1}}+\frac{\tld \Psi_{j+1}^{m}-\tld \Psi_{j}^{m}}{\Delta\tld y_{j+1}}-\frac{\tld \Psi_{j}^{m+1}-\tld \Psi_{j-1}^{m+1}}{\Delta\tld y_{j}}-\frac{\tld \Psi_{j}^{m}-\tld \Psi_{j-1}^{m}}{\Delta\tld y_{j}}\Big),
\end{align*}
\end{linenomath}
with $\tld t=m\Delta\tld t$ and an irregular grid with $\Delta\tld y_{j}$ the spacing between the node $j$ and $j-1$. Regrouping the unknowns at the next time step on the left hand side, the equation above can be expressed as the following linear system:
\begin{linenomath}
  \begin{equation}
    \tld\Psi^{m+1}_j = (\mat{P}_{lj})^{-1}(\tld S^m_l + \mat{Q}_{ln}\:\tld\Psi^{m}_n).
    \label{equ:tridiag}
  \end{equation}
\end{linenomath}
The matrix $\mat{P}_{lj}$ is tridiagonal and is efficiently inverted using Thomas algorithm.
\subsection{Elastodynamics}
The elastodynamics along the fault is solved through Equations (\ref{equ:elastodyn}) and (\ref{equ:convol}). Along a longitudinal grid of equally spaced sampling points $\tld x_i$, the non-local elastodynamic contribution is then integrated as
\begin{linenomath}
  \begin{equation}
    \tld \Phi^m_k = -k\tld q_0\tld\Omega^m_k + k\tld q_0\sum_{n=0}^{m} W_{{\rm III}, k}^{m-n}\dot{\tld\Omega}_k^n \Delta\tld t,
    \label{equ:dynastress}
  \end{equation}
\end{linenomath}
with $W_{{\rm III}, k}^{m}=W_{\rm III}(k\tld q_0m\Delta\t)$ being the convolution kernel whose integration according to Equation (\ref{equ:kernel}) is computed through a polynomials approximation \citep{morrissey97,abramowitz72}. $\tld \Omega_k$, $\dot{\tld \Omega}_k$ and $\tld\Phi_k$ are the discrete Fourier mode of respectively $\tld\delta_i$, $\tld V_i$ and $\tld\phi_i$. The discrete Fourier transform introduces a longitudinal periodic boundary condition with period $\tld X$ such that the fundamental wave number $\tld q_0=2\pi/\tld X$.
Combining Equations (\ref{equ:adim_elastodyn}) and (\ref{equ:vfromtau}), the continuity of shear stress through the gouge at time $\tld t=m\Delta\tld t$ requires that:
\begin{linenomath}
  \begin{equation}
\zeta(\tld\tau^m_i) \equiv \tld\tau^m_i - 1 +\sum_{j|\;y_j\in[-\frac{1}{2},\frac{1}{2}]}2e^{-\tld z}\sinh\Big(\frac{\tld\eta\tld z\tld\tau^m_i}{1-\tld p^m_{i,j}}\Big)\;\Delta\tld y - \tld{\phi}^m_i=0.
\label{equ:zeta}
\end{equation}
\end{linenomath}
$\zeta$ is a monotonic and differentiable function of $\tld\tau^m_i$ whose root is found iteratively using a Newton-Raphson scheme.
Finally, the slip velocity is integrated from the solution $\tld\tau^m_i$ as
\begin{linenomath}
  \begin{equation}
\tld V^{m}_i = \sum_{j|\;y_j\in[\frac{1}{2},\frac{1}{2}]}2e^{-\tld z}\sinh\Big(\frac{\tld\eta\tld z\tld\tau^m_i}{1-\tld p^m_{i,j}}\Big)\;\Delta\tld y.
\label{equ:integv}
\end{equation}
\end{linenomath}
\subsection{Coupled scheme}
At time $\tld t=m\Delta\tld t$, the coupled numerical scheme is integrated in time by the following predictor-corrector scheme:
\begin{enumerate}
    \item Compute $\tld p^{m+1}_j$ and $\tld T^{m+1}_j$ following Crank-Nicholson integration scheme (Equation \ref{equ:tridiag})
    \item Integrate predictor slip: $\tld\delta^*_i = \tld\delta^m_i + \Delta\tld t\; \tld V^{m}_i$
    \item Using the predictor slip, compute the predictor velocity $\tld V^*_i$ by solving Equations (\ref{equ:dynastress}-\ref{equ:zeta}-\ref{equ:integv})
    \item Correct slip integration: $\tld\delta^{m+1}_i = \tld\delta^m_i + \Delta\tld t(\tld V^{m}_i+\tld V^*_i)/2$
    \item Integrate the non-local dynamic contribution $\tld\phi^{m+1}_i$ using Equation (\ref{equ:dynastress})
    \item Compute frictional stress $\tld\tau^{m+1}_i$ using Equation (\ref{equ:zeta})
    \item Compute slip velocity $\tld V^{m+1}_i$ using Equation (\ref{equ:integv})
\end{enumerate}

\subsection{Initial and boundary conditions}

The system of interest consists of a shear zone lying between two semi-infinite linear elastic wall rocks as shown in Figure \ref{fig:system_setup}. The system is initially at rest under homogeneous pressure $p_0$, temperature $T_0$, shear $\tau_b$ and normal $\sigma_{\rm n}$ stress. At time $t=0$, the rupture event is nucleated by introducing a Gaussian perturbation of the pore pressure at the center of the fault such that
\begin{linenomath}
  \begin{equation}
    \tld p(\x,\y) = \tld \Pi \exp\Big(\frac{\tld x^2}{2\tld \xi^2}\Big), \;\;\; \mathrm{for}\;\y\in[-1/2,1/2].
  \end{equation}
\end{linenomath}
The amplitude and standard deviation are typically set respectively to $\tld\Pi=0.6$ and $\tld\xi=5\%\tld X$.
The system is assumed to be infinite in the transverse directions, whereas periodic boundary conditions in the longitudinal direction (with period $\tld X$) are assumed to compute the discrete Fourier modes $\tld \Omega_k$, $\dot{\tld \Omega}_k$ and $\tld\Phi_k$ used in Equation (\ref{equ:dynastress}).

\subsection{Stability and convergence}
Crank-Nicholson time integration defined in Equation (\ref{equ:tridiag}) is unconditionally stable and the definition of the stable time step is set by time integration of the elastodynamic model following the Courant-Friedrichs-Lewy condition:
\begin{linenomath}
  \begin{equation}
    \Delta \tld t = \beta_{\rm CFL}\Delta \tld x. 
    \label{equ:cfl}
  \end{equation}
\end{linenomath}
A value of $\beta_{\rm CFL}=0.1$ has been used in the simulations reported in this paper. The longitudinal direction is regularly discretized with $N_x$ sampling points, such that $\Delta\tld x = \tld X/N_x$. In the transverse direction, the gouge thickness is regularly sampled by $N_y+1$ points, such that $\Delta\tld y = 1/N_y$. To capture the exponential decay of pressure and temperature fields outside the gouge layer, the numerical domain symmetrically stretches into the surrounding wall rock, which is sampled with $N_y$ additional points that are logarithmically-spaced such that $\Delta\tld y_{j=1}=\Delta\tld y_{j=2N_y}=1$. Background pore pressure and temperature $\tld p = \tld T =0$ are imposed at the two edges ($y_{j=0}$ and $y_{j=2N_y}$) of the numerical grid, whose distance from the center of the gouge should exceed the diffusion length scale:
\begin{linenomath}
  \begin{equation}
    \tld y_{j=2N_y} > 2\sqrt{\tld\alpha N_t\Delta\t},
  \end{equation}
\end{linenomath}
with $\tld\alpha=\sqrt{\tld\alpha^2_{\rm hy}+\tld\alpha^2_{\rm th}}$ and $N_t$ being the total number of time steps.\\
Typical simulations use $N_x = 2048$ and $N_y = 200$, which are sufficient to reach numerical convergence.

\subsection{Rupture dynamics with weak strain localisation \label{sec:app_unconv}}

\begin{figure}
   \centering
   \includegraphics{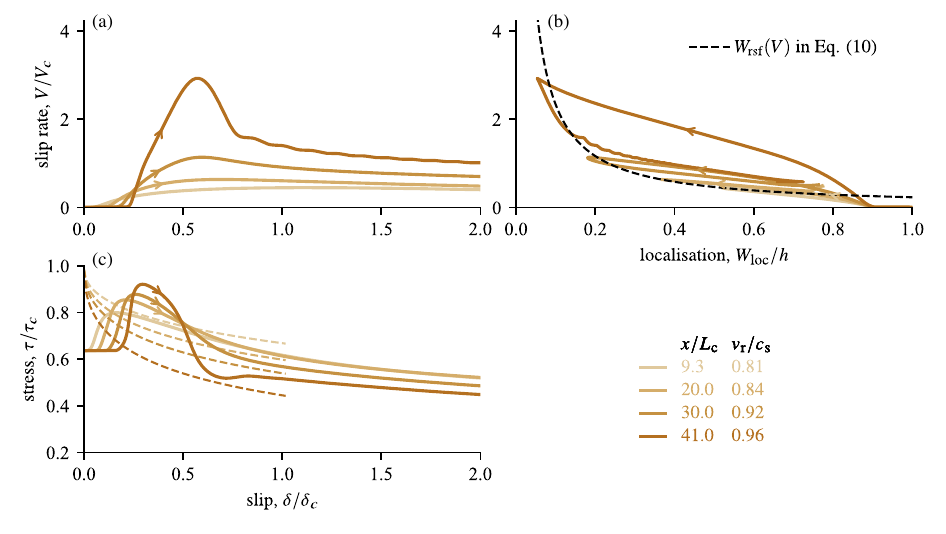}
   \caption{Same as Figure \ref{fig:space_snapshot} but for a simulation with four-times larger thermal and hydraulic diffusivities. Consequently, strain localisation remains very weak during the initial stage of the rupture when $v_{\rm r}<0.9c_{\rm s}$.}
   \label{fig:app_snapshot_undrained}
 \end{figure}

\begin{figure}
   \centering
   \includegraphics{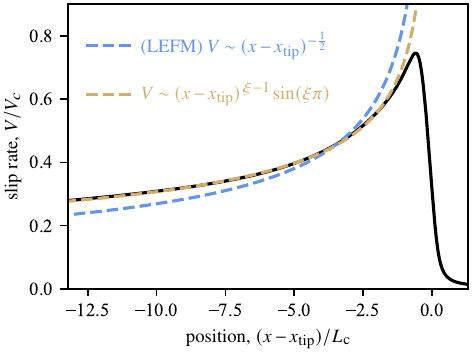}
   \caption{Snapshot of the slip velocity concentration near the rupture tip moving at $v_{\rm r}=0.85c_{\rm s}$ for the simulation with large thermal and hydraulic diffusivities and weak strain localisation shown in Figure \ref{fig:app_snapshot_undrained}. The near-tip singularity significantly deviates from LEFM prediction of Equation (\ref{equ:v_peak_pred}) but is well described by the asymptotic solution (\ref{equ:unconv_asymptot}) with an exponent $\xi=0.63$.}
   \label{fig:app_lefmfit_undrained}
 \end{figure}

The brown data set shown in Figures \ref{fig:stress_EBD}, \ref{fig:app_snapshot_undrained} and \ref{fig:app_lefmfit_undrained} correspond to a shear zone with (4$\times$) larger diffusivities, which lead to weaker strain localisation and larger localisation width following the predictions of Equations (\ref{equ:critwavelength}) and (\ref{equ:wloc}). As a result, the shear stress does not feature a sharp drop but rather a progressive weakening driven by the increase in pressure and temperature with no well-defined residual shear stress. Dynamic rupture can still be produce by using (4$\times$) larger domain and nucleation size but significantly differ from the dynamics of shear crack. In absence of a well-defined residual shear stress, the near-tip singular solution becomes \citep[e.g.][]{brener21}:
\begin{equation}
    V \sim (x-x_{\rm{tip}})^{\xi -1}\sin(\xi\pi),
    \label{equ:unconv_asymptot}
\end{equation}
with the singularity $\xi$ being one half at the vicinity of a standard LEFM shear crack (as in Equation \ref{equ:v_peak_pred}). Strong shear localisation produces a well-defined residual stress and a near tip singularity similar to standard shear crack, as demonstrated in the main text (Figure \ref{fig:lefm_fit}). This is no longer true for failure with weak strain localisation that features significantly different near-tip singularity and rupture dynamics (Figure \ref{fig:app_lefmfit_undrained}).

\end{document}